\documentclass{aa}
\def\a{\mbox{AE\,Aqr}\ }
\def\e{et~al.\ }

\def\r{R_{\rm m}}
\newcommand{\be}{\begin{equation}}
\newcommand{\ee}{\end{equation}}
\newcommand{\bdm}{\begin{displaymath}}
\newcommand{\edm}{\end{displaymath}}

\usepackage{graphicx}

\begin{document}

   \title{Can the 33\,s pulsations observed from AE~Aquarii be
explained in terms of accretion onto the white dwarf surface\,? }

   \author{Nazar Ikhsanov\inst{1,2}}

  \offprints{N.R.~Ikhsanov \\ \email{ikhsanov@mpifr-bonn.mpg.de}}

   \institute{Max-Planck-Institut f\"ur Radioastronomie, Auf dem
              H\"ugel 69, D-53121 Bonn, Germany
              \and
              Central Astronomical Observatory of the Russian
              Academy of Science at Pulkovo, Pulkovo 65--1, 196140
              Saint-Petersburg, Russia}

   \date{Received 18 April 2001 / Accepted 25 May 2001}

\authorrunning{Ikhsanov N.R.}
\titlerunning{Can the 33\,s pulsations be explained as an accretion
onto the white dwarf surface\,?}

\abstract{The 33\,s pulsing component observed from \a
is frequently assumed to be the result of accretion onto the
surface of a rotating white dwarf.
The validity of this assumption is discussed in the present paper.
I show that under the conditions of interest the white dwarf is in
the state of {\it supersonic propeller} and the efficiency of plasma
penetration into its magnetosphere is $\la 0.1\%$. This is too small
to explain the observed luminosity of the pulsing component.
Moreover, I find that for the currently established value of the
angle between the magnetic and rotational axes of the white dwarf,
the material entering the magnetosphere at the boundary can never 
reach its surface by flowing along the magnetic field lines. I
conclude  that the assumption about the direct accretion onto the
surface of  the white dwarf in \a contradicts the observational
data obtained in the optical/UV and thus cannot be accepted.
}

   \maketitle

   \section{Introduction}

AE~Aquarii is a non-eclipsing close binary system with an
orbital period $P_{\rm orb} \approx 9.88$\,hr and orbital
eccentricity $e \approx 0.02$. It is situated at the distance of
$\sim (100\pm 30)$\,pc. The normal component (secondary) is a K3-K5
main sequence red dwarf. The primary is a magnetized white dwarf
rotating with a period of 33\,s. The inclination angle of the system
is limited to $50^{\degr} < i < 70^{\degr}$ and the mass ratio
is $q=M_2/M_1=0.77\pm 0.03$ (for references see Table~1 in Ikhsanov
\cite{i00}).

\a emits detectable radiation in almost all parts of the spectrum.
It was observed as a powerful non-thermal flaring source in
the radio (Bastian \e \cite{bdch88}) and gamma-rays (Bowden \e
\cite{bbch92}; Meintjes \e \cite{mjr94}). The optical, UV and X-ray
radiation of the system is predominantly thermal and is coming from
at least three different sources. The visual light is dominated (up
to 95\%) by the secondary (Bruch \cite{bru91}; Welsh \e
\cite{welsh95}). The contribution of the primary is observed mainly
in the form of 33\,s (and 16.5\,s) coherent oscillations detectable in
the optical, UV and X-rays (Patterson \cite{p79}; Patterson \e
\cite{pbchr80}). Study of the optical-UV spectrum and profiles of the
pulsations (Eracleous \e \cite{erac94}) has shown that the pulsing
radiation is coming from two hot ($T \simeq 26000$\,K) spots with a
projected area of $\sim 4\,10^{16}\,{\rm cm^2}$, which are situated on
the white dwarf surface. Placing these spots at the magnetic pole
regions of the primary, Eracleous \e (\cite{erac94}) have estimated
the value of the angle between the rotational and magnetic axes of the
white dwarf to be $75^{\degr} \la \beta \la 77^{\degr}$. The remaining
light comes from a highly variable extended  source which manifests
itself in the blue/UV continuum, the optical/UV broad single-peaked
emission lines and the non-pulsing X-ray component. This source is
spread around the magnetosphere of the primary and is associated with
the mass-exchange process in the system (for a detailed system
description see Eracleous \& Horne \cite{eh96} and Ikhsanov
\cite{i00}).

On the grounds of the properties observed in the optical, \a was
classified as a nova-like star belonging to the DQ~Herculis subclass
of magnetic cataclysmic variables. Following this notion, it has been
suggested that the primary is a magnetized accretion-powered white
dwarf undergoing disk accretion (Patterson \cite{p79}).
Correspondingly, the flaring activity of the system was interpreted
in terms of a variable mass accretion rate onto the primary surface
(van Paradijs \e \cite{pka89}).

However extensive investigations of the system over the last five
years have clearly shown that \a does not fit this model. First,
studies of the 33\,s pulsations in the optical/UV (Eracleous \e
\cite{erac94}) and X-rays (Reinsch \e \cite{rbht95}; Clayton \&
Osborne \cite{co95}; Choi \e \cite{cda99}) have shown no correlation
between their amplitudes and the flaring of the system. This allowed
Eracleous \e (\cite{erac94}) to conclude that flares are not related
to the process of depositing material onto the primary surface.
Second, analysis of the H$\alpha$ Doppler tomogram of \a has shown
no evidence of a developed Keplerian accretion disk in the system.
Instead, it has shown that the material inflowing through the L1 point
into the Roche lobe of the primary is then streaming out from the
system with an average velocity $\sim 300\,{\rm km\,s^{-1}}$
(Wynn \e \cite{wkh97}; Welsh \e \cite{whg98}). Finally, de~Jager \e
(\cite{jmor94}) have reported the mean spindown rate of the white
dwarf $\dot{P}_{\rm s} = 5.64 \times 10^{-14}\,{\rm s\,s^{-1}}$ which
implies the spindown power of $L_{\rm sd}= I \Omega \dot{\Omega} \sim
10^{34}\,{\rm erg\,s^{-1}}$. $L_{\rm sd}$ exceeds the observed UV and
X-ray luminosities of the system by a factor of 120 and its
bolometric luminosity by a factor of more than
five\footnote{Hereafter the distance to \a is adopted to be 100\,pc}.
This indicates that the spindown power dominates the energy budget of
the system and raises the question of the nature of the spindown
torque, which would appear to be much larger than any inferred
accretion torque.

Four alternative answers to this question and, correspondingly,
four alternative theoretical models of the system are currently
under discussion: (i)~the magnetic propeller model (Wynn \e
\cite{wkh97}), (ii)~the spin-powered white dwarf pulsar model
(Ikhsanov \cite{i98}), (iii)~the gravitational wave emitter model
(Choi \& Yi \cite{cy00}) and (iv)~the differentially rotating white
dwarf model (Geroyannis \cite{g01}). Though these models essentially
differ from each other in many aspects they have an important common
point. Namely, the mass-exchange picture in \a is interpreted in
terms of the propeller action by the white dwarf. This implies that
the magnetic field of the white dwarf is strong enough for the
magnetospheric radius,    \be\label{rm}
\r = \eta \left(\frac{\mu^2}{\dot{M}\sqrt{2GM_1}}\right)^{2/7},
   \ee
to be larger than its corotational radius,
   \be\label{rcor}
R_{\rm cor} = 1.4\,10^9\ M_{0.8}^{1/3}\ P_{33}^{2/3}\,{\rm cm}.
     \ee
Here $\mu$, $M_{0.8}=M_1/0.8M_{\sun}$ and $P_{33}=P_{\rm s}/33$\,s are
the magnetic dipole moment, the mass and the spin period of the white
dwarf, respectively. $\dot{M}$ is the mass accretion rate onto the
primary magnetosphere and $0.5 \la \eta \la 1$ is the parameter which
accounts for the geometry of the accretion flow. Under these
conditions the white dwarf is in the {\it centrifugal inhibition
regime} and is able to eject the material out of the system due
to the propeller action.

One of the important questions, which remains to be explained
within this approach, is the origin of the hot spots on the white
dwarf surface. According to the canonical propeller model (e.g.
Davies \e \cite{dfp79}) no accretion occurs onto the surface of a
star in the state of propeller. This
indicates that the direct accretion onto the surface of the white
dwarf in \a cannot be responsible for the origin of the hot polar caps
and hence an alternative mechanism should be invoked. Following this
conclusion, it has been suggested that the polar caps are heated due
to dissipation of the magnetic or/and rotational energy of the
primary. In particular, due to the dissipation of Alfv\'en waves (Wynn
\e \cite{wkh97}) or/and due to the impact of the backflowing
relativistic particles accelerated in the magnetosphere of the white
dwarf (Ikhsanov \cite{i98}).

A different possibility to interpret the hot polar caps phenomenon
has been discussed by Kuijpers \e (\cite{kfa97}), Meintjes \&
de~Jager (\cite{mdj00}) and  Choi \& Yi (\cite{cy00}). These authors
pointed out that the mass accretion rate onto the white dwarf surface
required to explain the observed luminosity of the pulsing component
is $\dot{M}_{\rm a} \la 10^{14}\,{\rm g\,s^{-1}}$. This is essentially
smaller than the mass loss rate of the secondary 
($\dot{M} \sim 10^{16}\,{\rm g\,s^{-1}}$) derived from the observations 
of the Balmer continuum and the optical/UV emission lines
(see e.g. Eracleous \& Horne \cite{eh96}). On this basis they have
envisaged a situation in which the major fraction of the plasma
inflowing into the Roche lobe of the primary within the orbital plane
(so called low-altitude accretion flow) is ejected from the system
due to the propeller action, while a small amount of material, which
reaches the primary magnetosphere at the bases of the corotational
cylinder (so called high-altitude accretion flow)\footnote{According
to Patterson (\cite{p79}) and Eracleous \e (\cite{erac94}) the
rotational axis of the white dwarf in \a is almost perpendicular to
the orbital plane}, is able to penetrate through the magnetospheric
boundary and reach the white dwarf surface. The theoretical
analysis of this situation (which has not been performed so far) is
the subject of the present paper. I discuss the 
possible origin of the high-altitude accretion flow (Sect.\,2), 
the efficiency of plasma penetration into the primary magnetosphere
through the bases of the corotational cylinder (Sect.\,3) 
and the trajectory of the plasma inside the white dwarf
magnetosphere (Sect.\,4).  My basic conclusion is
that the assumption about the accretion nature of the 33\,s pulsing
component in \a has no sufficient theoretical grounds and thus, the
investigation of alternative mechanisms of the polar caps heating
seems to be more fruitful (Sect.~5).

    \section{The origin of high-altitude accretion flow}

In the general case the mass-exchange in a close binary system
can be realized due to the stream-fed and/or the wind-fed mass
transfer mechanisms. As shown by Ikhsanov (\cite{i97}) the mass
capture rate by the primary from the stellar wind of the normal
companion in \a is limited {\bf to} $\dot{M}_{\rm wa} \la
3.3\,10^{12}\,{\rm g\,s^{-1}}$. This value is significantly smaller
than that required to explain the observed UV/X-ray luminosity of the
pulsing component in terms of accretion onto the primary surface.
That is why the wind-fed mass transfer mechanism cannot be
responsible for the powerful high-altitude accretion flow in AE~Aqr.

The effective cross-section of the stream flowing into the Roche
lobe of the primary through the L1 point can be limited by the
cross-section of the throat at this point, which in the case of
\a is (Ikhsanov \cite{i97})
   \be\label{cros}
Q = \frac{2 \pi c_{\rm s}^{2} a^{3}}{k G (M_1 + M_2)} \sim 1.85
\times 10^{19} \left(\frac{T}{10^{4}\,{\rm K}}\right) \ {\rm cm^{2}}.
      \ee
Here $c_{\rm s}$ and $T$ are the sound speed and the plasma
temperature at the L1 point, $a$ is the orbital
separation, $M_1$ and $M_2$ are the masses of the primary and the
secondary, respectively, and $k$ is a dimensionless constant
depending on the mass ratio of the components (for discussion see
Meyer \& Meyer-Hofmeister \cite{mmh83}). This allows to evaluate
the effective radius of the stream at the L1 point\footnote{Here I
would like to note that the plasma density across the stream at the L1
point is $n_{\rm s} \propto \exp\{-(r/r_{\rm s0})^2\}$.} as $r_{\rm
s}=\sqrt{Q\pi^{-1}}\simeq 2.4\,10^9$\,cm.

The stream inside the Roche lobe of the primary follows the ballistic
trajectory toward the white dwarf until it interacts with the star
magnetic field. The maximum possible expansion of the stream during
this part of its trajectory is
  \be
\Delta r \la t_{\rm ff} c_{\rm s} \simeq 2\,10^9\
R_{11}^{3/2}\ M_{0.8}^{-1/2}\ T_4^{1/2}\,{\rm cm},
  \ee
where $t_{\rm ff}=R^{3/2}/\sqrt{2GM_1}$ is the free-fall time,
$R_{11}$ is the distance from the white dwarf to the L1
point expressed in units of $10^{11}$\,cm and
$T_4=T/10^4$\,K. Hence the altitudes at which the stream flows into
the white dwarf magnetosphere can be estimated as
   \be
\mid\kappa\mid \la \arcsin{\left(\frac{r_{\rm s}+\Delta r}{R_{\rm
L1}}\right)} \simeq 3^{\degr},
   \ee
i.e. the material inflowing through the L1 point reaches the white
dwarf magnetosphere in a narrow band around the rotational equator of
the primary.

The situation could be different if the secondary in \a essentially
overflows its Roche lobe. This assumption, however, contradicts the
results of the investigation of gradual variations of optical
brightness in the quiescent state (van Paradijs \e \cite{pka89}). 
Furthermore, in this case, the mass transfer rate in the system 
would be essentially higher than that currently estimated (see
Introduction).

An additional possibility to explain the origin of the high-altitude
accretion flow is to assume that the temperature of the material
surrounding the stream at the L1 point is about of $10^7$\,K. In
this case the value of $\Delta r$ proves to be comparable to the
magnetospheric radius of the white dwarf and, hence, the spherically
symmetrical accretion flow onto the primary magnetosphere can be
expected. In principle this assumption could be accepted if the
mechanism of plasma heating at the L1 point is explained.

    \section{The efficiency of plasma entry into the magnetosphere
of the white dwarf}

As shown by Arons \& Lea (\cite{al76}) the mass accretion rate onto
the surface of a magnetized compact star is limited by the rate of
plasma penetration into the star magnetosphere. The latter depends on
the shape of the magnetospheric boundary, the geometry of
the accretion flow and the physical parameters of plasma over the
boundary. Correspondingly, three modes of plasma penetration into the
star magnetic field can be realized: the interchange instabilities,
the diffusion, and the reconnection of the magnetic field lines (for
discussion see Elsner \& Lamb \cite{el84}).

Choi \& Yi (\cite{cy00}) have distinguished two components of the
accretion flow beyond the white dwarf magnetosphere: 
(i)~the accretion stream (the low-altitude component)
and (ii)~the almost spherically symmetrical accretion flow (the
high-altitude component).
As shown by Wynn \e (\cite{wkh97}) the stream interacts with the
magnetic field of the white dwarf in a local region at low altitudes 
($\mid\kappa\mid \la 3{\degr}$) mainly at the closest approach to 
the primary ($r_{\rm ms} \simeq 10^{10}$\,cm). 
The influence of the stream on the large scale
field of the white dwarf at higher altitudes (i.e. $\mid\kappa\mid
\gg 3^{\degr}$) is small and the dipole approximation for the field 
of the primary in these regions can be used.

The interaction between the spherically symmetrical accretion flow
and the dipole magnetic field of a compact star in the state of
propeller has been investigated by Davies \e (\cite{dfp79}) and
Davies \& Pringle (\cite{dp81}). They have shown that the
magnetosphere of the primary in this case is closed\footnote{The
shape of the magnetosphere is similar to that derived by Arons
\& Lea (\cite{al76})} and the magnetospheric radius can be evaluated as
     \be
r_{\rm c} \ga \r = 1.8\,10^{10} \mu_{32}^{4/7}
\dot{M}_{15}^{-2/7} M_{0.8}^{-1/7}\,{\rm cm},
  \ee
where $\mu_{32}=\mu/10^{32}\,{\rm G\,cm^3}$ and $\dot{M}_{15}$ is 
the mass accretion rate of the high-altitude component onto the 
magnetospheric boundary expressed in units of 
$10^{15}\,{\rm g\,s^{-1}}$.

Under the condition $r_{\rm c} > R_{\rm cor}$, that is just the
case of the white dwarf in AE~Aqr, the state of the primary is
classified as {\it supersonic propeller} (the linear velocity of the
magnetospheric boundary, $v_{\varphi}=2\pi r_{\rm c}/P_{\rm s}$,
exceeds the sound speed in the accretion flow,
which is limited by the free-fall velocity, $V_{\rm
ff}=\sqrt{2GM_1/r_{\rm c}}$). In this situation the magnetosphere 
of the
star is surrounded by a turbulent atmosphere in which the plasma
temperature is of the order of the free-fall temperature $T_{\rm
ff}=GMm_{\rm p}/kr_{\rm c}$ (where $m_{\rm p}$ and $k$ are the proton
mass and the Boltzmann constant, respectively).

As shown by Davies \& Pringle (\cite{dp81}) the rate of plasma
penetration through the magnetic field of a compact star, which is in
the state of the supersonic propeller, is almost negligible. In the
region with $r_{\rm c} \cos{\kappa} > R_{\rm cor}$ (which in our case
corresponds to the magnetic latitudes $\mid\kappa\mid \la 85^{\degr}$)
the centrifugal force dominates the gravitational force at the
magnetospheric boundary. That is why the plasma entering the
magnetosphere in this region is unable to flow along the field
lines to the primary surface but is pushed back by the fast rotating
magnetosphere.  In the region $r_{\rm c} \cos{\kappa} \la R_{\rm
cor}$ (which corresponds to $85^{\degr} \la \mid\kappa\mid \la
90^{\degr}$) the centrifugal force is not effective. However, the
effective penetration of plasma into the magnetosphere does not occur
since the temperature of the plasma over the boundary is  $T(r_{\rm
c}) \sim T_{\rm ff}$ and hence, the magnetospheric boundary is stable
with respect to interchange instabilities (see Arons \& Lea
\cite{al76}; Elsner \& Lamb \cite{el77}). In this situation the
efficiency of plasma penetration into the magnetic field of the
primary (due to the magnetic field lines reconnection and/or
diffusion) is smaller than 1\% (see Ikhsanov \cite{i01} and
references therein). Furthermore, the area of the bases of the
corotational cylinder in the considered case constitutes about 10\%
of the total magnetospheric area. That is why the rate of plasma
penetration into the white dwarf magnetosphere is at least three
orders of magnitude smaller than the initial mass accretion rate in
the high-altitude accretion flow. This, however, is too small to
explain the luminosity of the 33\,s pulsing component.

    \section{Accretion flow inside the primary magnetosphere}

The last item I briefly address in this section is the plasma
flow inside the white dwarf magnetosphere. It is well known that
the accreting plasma penetrates the magnetosphere across the
filed lines forming the magnetopause, which is situated at the
boundary and has the thickness $\delta_{\rm m} \ll \r$ (see e.g.
Arons \& Lea \cite{al76}, Ghosh \& Lamb \cite{gl79} and Elsner \&
Lamb \cite{el84}). Plasma penetrating the magnetosphere
flows along the field lines within a relatively narrow channel
towards the star surface. Following this way it can reach the
star surface if all parts of the channel are situated inside the
corotational cylinder (otherwise it will be pushed out from the
magnetosphere by the centrifugal force). This condition is always
satisfied if the magnetospheric radius of the star is smaller than
the corotational radius, i.e. if the star is in the accretor state.
But if the star is in the propeller state the realization of this
condition is not obvious. The answer depends on whether a magnetic
line, which is situated inside the corotational cylinder and
connects the star surface with the bases of the corotational
cylinder, exists in the magnetosphere.

To investigate this problem I use the dipole approximation for the
magnetic field of the primary\footnote{This approximation is valid in
the region $R_{\rm wd} < r < r_{\rm c}$ (see Arons \& Lea
\cite{al76}).}. A magnetic line of force in this case has the
equation
    \be\label{1}
r=r_{\rm e} \cos^2{\lambda},
  \ee
where $\lambda$ is the magnetic latitude and $r_{\rm e}$ is the
distance from the line origin to the point of its intersection with
the plane of the magnetic equator\footnote{In the azimuthal direction
the equation of the magnetic line of force is $\varphi=$\,const.}. I
denote the angle between the magnetic and rotational axes of the
primary by $\beta$ and the angle between $\vec{B}$ and the radius
vector by $\alpha$. The latter can be expressed, according to Alfv\'en
\& F\"althammar (\cite{af63}), as
   \be\label{alpha}
\sin{\alpha} = \frac{\cos{\lambda}}{\sqrt{1+3 \sin^2{\lambda}}}.
  \ee

\begin{figure}
\resizebox{\hsize}{!}{\includegraphics{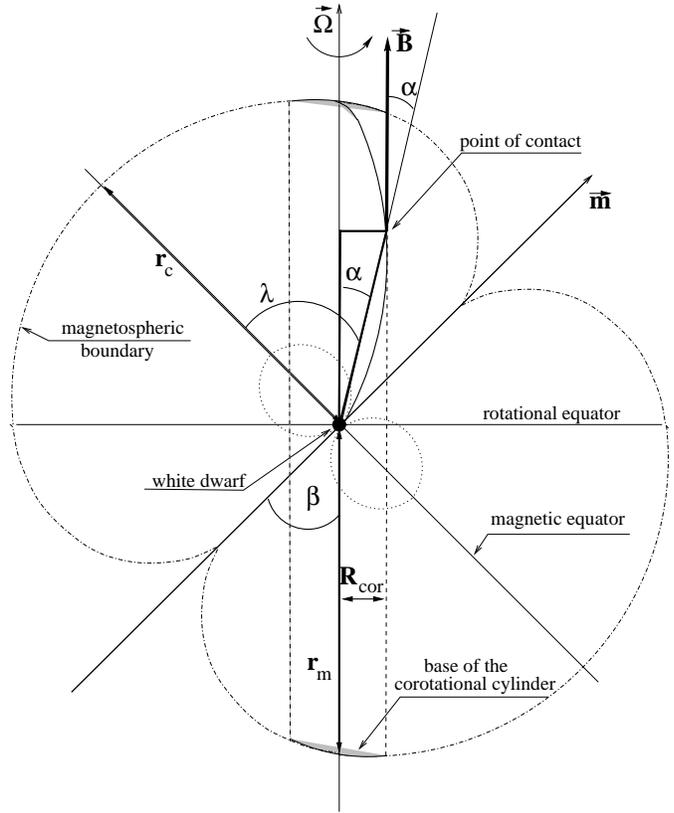}}
\caption{Schematic description of the magnetosphere of the white
dwarf undergoing high-altitude accretion as discussed in the
text}
\label{F1}
\end{figure}

A situation in which at least one line of force is situated inside
the corotational cylinder and connects the star surface with the
magnetospheric boundary is shown in Fig.~\ref{F1}. This line has a
point of contact with the surface of the corotational cylinder in
which the vector $\vec{B}$ is parallel to the rotational axis of the
star, $\vec{\Omega}$. As it is seen from the figure the following
conditions are satisfied in this case 
   \be\label{beta}
\beta=\left(\frac{\pi}{2} - \lambda\right) + \alpha,
  \ee
  \be\label{2}
\frac{R_{\rm cor}}{r_{\rm e} \cos^2{\lambda}} = \sin{\alpha}.
  \ee
The value $r_{\rm e}$ for this line can be evaluated taking into
account that it reaches the magnetospheric boundary at $\lambda =
(\pi/2 - \beta)$. Hence,
  \be
r_{\rm e} = \frac{r_{\rm m}(\lambda)}{\cos^2{(\pi/2-\beta)}},
  \ee
where $r_{\rm m}(\lambda)$ is the distance from the white dwarf to the
base of the corotational cylinder. According to Arons \& Lea
(\cite{al76}) in the case of spherically symmetric accretion,
$r_{\rm m}(\lambda)$  can be approximated as
   \be\label{rmb}
\frac{r_{\rm m}}{r_{\rm c}} \simeq\left\{
\begin{array}{lc}
(\cos{\lambda})^{0.2639}, \hspace{0mm}& {\rm for}
\hspace{2mm} \mid\lambda\mid \la \lambda_0, \\
& \\
0.51 + 0.63\left(1-\frac{2\mid\lambda\mid}{\pi}\right)^{2/3},
\hspace{0mm}& {\rm for} \hspace{2mm} \mid\lambda\mid > \lambda_0,
  \end{array}
   \right.
    \ee
where $\lambda_0 \simeq 80^{\degr}$.

Combining Eqs.~(\ref{1}--\ref{rmb}) I find that plasma accretion onto
the primary surface in the considered case could be realized if the
following condition is satisfied
   \be
\frac{\tan{\lambda} \sin{\lambda}}{\sqrt{1+3 \sin^2{\lambda}}}\ -\
\frac{r_{\rm c}}{9 R_{\rm cor}} (\cos{\lambda})^{0.2639}\ \ga\  0.
  \ee
Solving this inequality for $\lambda$ and putting the result to
Eqs.~(\ref{alpha}) and (\ref{beta}) I find that the accretion process
onto the surface of the white dwarf in \a could be
realized\footnote{The corotational radius and the equatorial
magnetospheric radius of the white dwarf are taken as $R_{\rm cor}=
1.5\,10^9$\,cm and $r_{\rm c} = 1.8\,10^{10}$\,cm, respectively (see
above)} if $\beta \la 38^{\degr}$. At the same time, the value of
$\beta$ evaluated by Eracleous \e (\cite{erac94}) from the
investigation of the pulse profiles of the 33\,s oscillations is
$75^{\degr} \la \beta \la 77^{\degr}$. This indicates that the
assumption about the direct accretion of plasma onto the surface of
the white dwarf in \a contradicts the results of optical/UV
observations of the system and thus cannot be accepted.

    \section{Conclusions}

One can conclude that the assumption about
the direct plasma accretion onto the surface of the white dwarf in \a
cannot be accepted. The theoretical grounds of this conclusion are
the following. First, the origin of the  high-altitude accretion
flow, which moves directly to the bases of the corotational cylinder
of the white dwarf with the rate $> 10^{13}\,{\rm g\,s^{-1}}$,
in the particular case of AE~Aqr, is rather unclear. Second, even if
we assume that this accretion flow exists, the efficiency of its
penetration into the white dwarf magnetosphere is smaller than
0.1\%. Finally, for the established value of the angle between the
rotational and magnetic axes of the white dwarf in AE~Aqr, the
material penetrating the magnetosphere through the bases of
the corotational cylinder could never reach the star surface
flowing along the magnetic field lines.

From the observational point of view the assumption about the
accretion onto the surface of the white dwarf in \a also faces a
number of serious problems. In particular, the temperature of
the plasma in the polar caps of the white dwarf in \a ($T\simeq
26000$\,K) derived by Eracleous \e (\cite{erac94}) is significantly
smaller than the surface temperature in the magnetic pole regions of
accreting white dwarfs (typically $T\sim 10^8$\,K, Eracleous \e
\cite{ehp91}). Furthermore, the X-ray spectrum of \a is soft and
essentially differs from the hard X-ray spectra of all intermediate
polars as well as from those of almost all accretion powered close
binaries (Clayton \& Osborne \cite{co95}). On the other hand, due to
the power law spectrum with $\alpha \approx -2$ and the ratio of
the X-ray luminosity to the spindown luminosity $L_{\rm x}/L_{\rm sd}
\sim 10^{-3}$, the X-ray emission of \a is rather similar to the
X-rays detected from spin-powered pulsars (Becker \& Tr\"umper
\cite{bt97}). This resemblance indicates that the energy release in
the magnetosphere of the white dwarf in \a and the magnetospheres of
spin-powered pulsars may have a common nature and, if so, the
assumption about the accretion onto the white dwarf surface proves
not to be necessary.

\begin{acknowledgements}
I would like to thank the referee, Prof. Antonio Bianchini, for
careful reading the manuscript and suggested improvements.
I acknowledge the support of the Follow-up program of the Alexander
von Humboldt Foundation.
\end{acknowledgements}

\end{document}